\documentclass[aps,twocolumn,showpacs,preprintnumbers,amsmath,amssymb,showpacs,showkeys]{revtex4}
\usepackage{amssymb}
\usepackage{amsmath}
\usepackage{amsfonts}
\usepackage{epsfig}
\usepackage{graphicx}
\usepackage{tabularx}
\newcommand{\seq}{\begin{subequations}}
\newcommand{\sen}{\end{subequations}}
\newcommand{\eq}{\begin{eqnarray}}
\newcommand{\en}{\end{eqnarray}}

\def\shiftdown#1{#1\llap{\lower.04ex\hbox{#1}}}

\newcommand{\ra}{\rangle}
\newcommand{\la}{\langle}

\newcommand{\bfq}{{\bf q}_{\perp}}

\newcommand{\bfk}{{\bf k}_{\perp}}

\begin{document}

\title{Light-front quark model consistent with \\
Drell-Yan-West duality and quark counting rules}

\author{
Thomas Gutsche$^1$,
Valery E. Lyubovitskij$^1$ 
\footnote{On leave of absence 
from Department of Physics, Tomsk State University,  
634050 Tomsk, Russia},
Ivan Schmidt$^2$, 
Alfredo Vega$^3$ 
\vspace*{.6\baselineskip}\\
}

\affiliation{
$^1$ Institut f\"ur Theoretische Physik,
Universit\"at T\"ubingen, \\
Kepler Center for Astro and Particle Physics,
\\ Auf der Morgenstelle 14, D-72076 T\"ubingen, Germany
\vspace*{.6\baselineskip} \\
\hspace*{-1cm}
$^2$ Departamento de F\'\i sica y Centro Cient\'\i
fico Tecnol\'ogico de Valpara\'\i so (CCTVal), Universidad T\'ecnica
Federico Santa Mar\'\i a, Casilla 110-V, Valpara\'\i so, Chile
\vspace*{.6\baselineskip} \\ 
$^3$ Departamento de F\'isica y Astronom\'ia, \\
Universidad de Valpara\'iso,\\
Avenida Gran Breta\~na 1111, Valpara\'iso, Chile
\vspace*{.5\baselineskip} \\ 
}

\date{\today}

\begin{abstract}

We propose a phenomenological light-front wave function for hadrons with  
arbitrary twist-dimension (mesons, baryons and multiquark states),  
which gives the correct scaling behavior of structure functions  
(parton distributions) and form factors for pions and nucleons.  
For other hadronic states the proposed wave function produces form factors  
consistent with quark counting rules,  
and gives predictions for structure functions (parton distributions).  
As an application we build a light-front quark model for nucleons based   
on the proposed wave function. 

\end{abstract}



\maketitle

\section{Introduction}

The main objective of this paper is to propose  
a phenomenological light-front wave function (LFWF) for hadrons with  
arbitrary twist-dimension (mesons, baryons and multiquark states)
which gives the correct scaling behavior of structure functions  
(parton distributions) and form factors both for pions and nucleons.  
For other hadronic states the proposed wave function produces form factors  
consistent with quark counting rules,  
and also gives predictions for the corresponding structure functions  
(parton distributions).  
As an application we construct a light-front quark model for nucleons  
based on the proposed wave function. This model  
is by construction consistent with the Drell-Yan-West (DYW) 
relation~\cite{Drell:1969km}  
between the large-$Q^2$ behavior of nucleon electromagnetic form factors and 
the large-$x$ behavior of the structure functions (see also  
Ref.~\cite{Bloom:1970xb} for the extension to inelastic scattering), and  
with quark counting rules~\cite{Brodsky:1973kr}. Based on the findings 
of Refs.~\cite{Drell:1969km,Bloom:1970xb,Brodsky:1973kr} one can e.g.  
relate the behavior of the quark distribution function in the nucleon  
$q_v(x) \sim (1-x)^p$ at $x \to 1$ to the scaling of the proton Dirac  
form factor $F_1^p(Q^2) \sim 1/(Q^2)^{(p+1)/2}$ at large $Q^2$
(the parameter $p$ is related to the number of valence constituents $N$ 
in the hadron, hence for $N=3$ we have $p=3$). 
In Refs.~\cite{Brodsky:1994kg,Yuan:2003fs}  
the large-$x$ scaling of pion and nucleon PDFs and GPDs has been obtained  
in the framework of perturbative QCD. In particular, the pion structure  
function behaves as $(1-x)^2$ at $x \to 1$, while nucleon spin non-flip 
and spin-flip structure functions behave at large $x$ as $(1-x)^3$ and 
$(1-x)^5$, respectively. The importance of these scaling laws and their role 
in the description of hadron structure has been stressed and studied 
in detail in the literature, see e.g. Refs.~\cite{Brodsky:1994kg}-\cite{Diehl:2004cx}. 

In this paper we show that these important features of nucleon structure  
can be also fulfilled in the framework of a simple light-front quark model  
based on a phenomenological wave function with a specific dependence on the  
transverse momentum $\bfk$ and the light-cone variable $x$. 
We also derive the phenomenological LFWF for hadrons with arbitrary 
twist-dimension $\tau$, which gives the required scaling for the
corresponding form factors and provides predictions 
for unknown structure functions and GPDs. 

\section{Light-front wave function for hadron with arbitrary twist} 

In this section we propose a phenomenological LFWF for hadrons with 
arbitrary twist-dimension $\tau$ without referring to the flavor 
structure. We choose the form 
\eq\label{Eq_LFWF}  
\psi_\tau(x,\bfk) &=& N_\tau \, \sqrt{\log(1/x)} \, 
(1-x)^{\frac{\tau}{2}+\frac{\alpha(\tau)}{2}-1} \nonumber\\[1mm] 
&\times& \exp\biggl[- \frac{\bfk^2}{2\kappa^2} 
\, \frac{\log(1/x)}{(1-x)^{2-\alpha(\tau)}} 
\biggr]
\en 
where 
\eq 
N_\tau &=& \frac{4\pi}{\kappa} \, \sqrt{1+\tau} 
\en 
is the normalization constant, 
\eq
\alpha(\tau) = \frac{2}{\tau-1} \,, 
\en 
$x$ is the Bjorken variable, 
$\bfk$ is the transverse momentum and $\kappa$ is the scale parameter. 
Such type of LFWF is motivated by the soft-wall AdS/QCD model 
(see e.g. Ref.~\cite{Brodsky:2007hb,Brodsky:2008pg,%
Vega:2009zb,Branz:2010ub,Brodsky:2011xx,Gutsche:2012ez}) and 
corresponds to the dressed function of a hadron dual to the corresponding 
bulk profile confined in the quadratic dilaton potential. 
The idea to extract LFWFs by matching to AdS/QCD has 
originally been suggested in Ref.~\cite{Brodsky:2007hb}, considering 
the pion electromagnetic form factor in two approaches -- AdS/QCD and 
light-front QCD. In a series of papers~\cite{Brodsky:2007hb,%
Brodsky:2008pg,Vega:2009zb,Branz:2010ub,Brodsky:2011xx,Gutsche:2012ez}  
this problem was further discussed in detail. 
The LFWF derived in the present paper is different from the LFWFs directly 
extracted from AdS/QCD. The difference lies in an additional specific dependence 
on the light-cone variable, which is required to get consistency with model-independent 
scaling rules. In particular, the effective LFWF extracted from AdS/QCD 
in Ref.~\cite{Brodsky:2011xx}, in the case of two-parton (meson) states, has 
the following form 
\eq 
\psi_{\tau=2}(x,\bfk) \sim \frac{\sqrt{\log(1/x)}}{1-x} 
\exp\biggl[- \frac{\bfk^2}{2\kappa^2} 
\frac{\log(1/x)}{(1-x)^2} \biggr] \,. 
\en 
The LFWF as proposed in (\ref{Eq_LFWF}) contains the extra factor 
\eq 
f_L(x) = (1-x)^2,
\en 
depending on the light-cone variable with the result
\eq 
\psi_{\tau=2}(x,\bfk) &\sim& \frac{\sqrt{\log(1/x)}}{1-x} f_L(x) 
\nonumber\\
&\times& \exp\biggl[- \frac{\bfk^2}{2\kappa^2} 
\frac{\log(1/x)}{(1-x)^2} f_L(x) \biggr] \,. 
\en  
The LFWF suggested here leads to the structure function (parton distribution) 
$\rho_\tau(x)$ 
\eq 
\rho_\tau(x) = \int\frac{d^2\bfk}{16\pi^3} |\psi_\tau(x,\bfk)|^2 
= (1 + \tau) (1-x)^\tau \,, 
\en 
which is normalized to 1 
\eq 
\int_0^1 dx \, \rho_\tau(x) = 1 
\en  
and gives the required behavior of parton densities at large $x \to 1$, 
both in the case of the pion $(1-x)^2$ and the nucleon $(1-x)^3$ (helicity 
non-flip density). The helicity-flip density for 
the nucleon will be discussed in Sec.III. 

Now we calculate the form factor $F_\tau(Q^2)$ and the 
generalized parton distribution (GPD) $H_\tau(x,Q^2)$ 
\eq 
F_\tau(Q^2) &=& \int\limits_0^1 dx H_\tau(x,Q^2) \nonumber\\
&=& \int\limits_0^1 dx \int\frac{d^2\bfk}{16\pi^3} 
\psi^\ast_\tau(x,\bfk^\prime) \, \psi_\tau(x,\bfk) \\
&=& \int\limits_0^1 dx \, \rho_\tau(x) \, 
\exp\biggl[- \frac{Q^2}{4\kappa^2} \, 
\log(1/x) \, (1-x)^{\alpha(\tau)}\biggr] \nonumber 
\en 
where $\bfk' = \bfk + (1-x) \bfq$ and $\bfq$ 
is the momentum transfer. 
One can see that the $Q^2$-dependence of the GPD generalizes 
the so-called modified Regge ansatz~\cite{Burkardt:2002hr} 
for the nucleon helicity non-flip GPD 
\eq 
{\cal H}(x,Q^2) \sim (1-x)^3 \, 
e^{- \frac{Q^2}{\Lambda^2} \log (1/x) (1-x)} 
\en 
for the specific choice $\tau = 3$ to the case of hadrons with arbitrary 
twist-dimension $\tau$, where $\Lambda$ is the scale parameter. 
E.g. for the pion we get  
\eq 
{\cal H}_\pi(x,Q^2) \sim (1-x)^2 
\, e^{- \frac{Q^2}{\Lambda^2} \log (1/x) (1-x)^2} \,.
\en 
It is easy to check that at large $Q^2$ the hadronic form factors 
are consistent with quark counting rules 
\eq 
F_\tau(Q^2) \sim \frac{1}{(Q^2)^{\tau - 1}} \,. 
\en 
Next, following the ideas of Refs.~\cite{Brodsky:2007hb,Brodsky:2008pg,%
Vega:2009zb,Branz:2010ub,Gutsche:2012ez}, we discuss the inclusion of 
quark masses in the proposed LFWF 
effective wave functions. We use a two-body approximation 
where $m_1$ and $m_2$ are treated as the masses 
of the struck quark and spectator compound, which is the spectator quark 
in the case of mesons, the diquark in the case of baryons, 
and a multiquark cluster in the case 
of other hadrons. Inclusion of quark masses leads to
\eq\label{Eff_WF} 
\psi_\tau(x,\bfk) &\to& \psi_\tau(x,\bfk,m_1,m_2) \nonumber\\[1mm]
&=& N_\tau \, \sqrt{\log(1/x)} \, x^{\beta_1} \, 
(1-x)^{\beta_2 + \frac{\tau}{2}+\frac{\alpha(\tau)}{2}-1} 
\nonumber\\[1mm]
&\times& \exp\biggl[- \frac{{\cal M}^2}{2\kappa^2} \, 
\frac{x \, \log(1/x)}{(1-x)^{1-\alpha(\tau)}} \biggr] \,, 
\en 
where 
\eq 
{\cal M}^2 = {\cal M}^2_0 + \frac{\bfk^2}{x (1-x)} 
 = \frac{\bfk^2 + m_1^2}{x} \, + \, \frac{\bfk^2 + m_2^2}{1-x} 
\en 
is the invariant mass. 
Here $\beta_1$ and $\beta_2$ are parameters, which depend on the masses $m_1$ 
and $m_2$, respectively. One of these parameters (the parameter $\beta_2$) 
must obey the constraint $\beta_2 \to 0$ in the chiral limit 
$m_1, m_2 \to 0$ in order to have consistency with scaling rules.

\section{Light-front quark-diquark model for the nucleon} 

In this section we consider the application of the phenomenological LFWF 
to the derivation of a light-front quark-diquark model for the nucleon. 
First we collect the well-known relations~\cite{Radyushkin:1998rt} 
between the set of nucleon Dirac and Pauli 
form factors $F_{1,2}^N$ ($N=p,n$) to the ones defining 
the distribution of valence quarks in nucleons $F_{1,2}^q$ ($q=u,d$) 
and the valence quark generalized parton distributions 
(GPDs)~\cite{Mueller:1998fv} ($\mathcal{H}^{q}$ and 
$\mathcal{E}^{q}$)
\eq\label{FF}
F_i^{p(n)}(Q^2) &=& \frac{2}{3} F_i^{u(d)}(Q^2) 
                 -  \frac{1}{3} F_i^{d(u)}(Q^2)\,, 
\nonumber\\
F_1^q(Q^2) &=& \int_{0}^{1} dx \, \mathcal{H}^{q}(x,Q^2)  \,,\\
F_2^q(Q^2) &=& \int_{0}^{1} dx \, \mathcal{E}^{q}(x,Q^2)  \, .
\nonumber 
\en 
At $Q^2=0$ the GPDs are related to 
the quark densities --- valence $q_v(x)$ and magnetic $\mathcal{E}_{q}(x)$ 
\eq\label{CondicionDensidadQuark}
\mathcal{H}^{q}(x,0)=q_{v}(x)\,, 
\quad 
\mathcal{E}^{q}(x,0)=\mathcal{E}^{q}(x) \,, 
\en
which are normalized as 
\eq\label{normalization} 
n_q &=& F_1^q(0) = \int\limits_0^1 dx \, q_v(x)\,, \nonumber\\
\kappa_q &=& F_2^q(0) = \int\limits_0^{1} dx \, \mathcal{E}^q(x)\, . 
\en
The number of $u$ or $d$ valence quarks in the proton is denoted
by $n_q$ and $\kappa_q$ is the quark anomalous 
magnetic moment. 

Next we recall the definitions of the nucleon Sachs form factors 
$G_{E/M}(Q^2)$ and the electromagnetic $\la r^2_{E/M} \ra^N$ radii 
in terms of the Dirac and Pauli form factors $F_{1,2}^N(Q^2)$  
\eq 
G_E^N(Q^2) &=& F_1^N(Q^2) - \frac{Q^2}{4m_N^2} F_2^N(Q^2)\,, 
\nonumber\\ 
G_M^N(Q^2) &=& F_1^N(Q^2) + F_2^N(Q^2)\,, \nonumber\\
\la r^2_E \ra^N &=& - 6 \frac{dG_N^E(Q^2)}{dQ^2}\bigg|_{Q^2 = 0} \,, 
\nonumber\\
\la r^2_M \ra^N &=& - \frac{6}{G_M^N(0)} \, 
\frac{dG_M^N(Q^2)}{dQ^2}\bigg|_{Q^2 = 0}  \,, 
\en 
where $G_M^N(0) \equiv \mu_N$ is the nucleon magnetic moment. 

The light-front representation~\cite{Brodsky_Drell,LFQCD}  
for the Dirac and Pauli quark form factors is  
\eq\label{FF_LFQCD} 
F_1^q(Q^2) &=& 
\int\limits_0^1 dx 
\int\frac{d^2\bfk}{16\pi^3} \, 
\biggl[ 
\psi_{+q}^{+\, \ast}(x,\bfk')\psi_{+q}^+(x,\bfk) 
\nonumber\\
&+&\psi_{-q}^{+\, \ast}(x,\bfk')\psi_{-q}^+(x,\bfk) 
\biggr] \,, \nonumber\\
& &\\
F_2^q(Q^2) &=& - \frac{2M_N}{q^1-iq^2}
\int\limits_0^1 dx 
\int\frac{d^2\bfk}{16\pi^3} \nonumber\\
&\times&
\biggl[ 
\psi_{+q}^{+\, \ast}(x,\bfk')\psi_{+q}^-(x,\bfk) 
\nonumber\\
&+&\psi_{-q}^{+\, \ast}(x,\bfk')\psi_{-q}^-(x,\bfk) 
\biggr] \,. \nonumber 
\en  
Here $M_N$ is the nucleon mass, $\psi_{\lambda_q q}^{\lambda_N}(x,\bfk)$ 
are the LFWFs with specific helicities of the nucleon $\lambda_N  = \pm$ 
and the struck quark $\lambda_q = \pm $, where plus and minus correspond to 
$+\frac{1}{2}$ and $-\frac{1}{2}$, respectively. 
We work in the frame with $q=(0,0,\bfq)$, and therefore 
the Euclidean momentum squared is $Q^2 = \bfq^2$. 

In the quark-scalar diquark model the generic ansatz for the LFWFs reads 
\eq 
\psi_{+q}^+(x,\bfk) &=& \frac{m_q + xM_N}{x} \, \varphi_q(x,\bfk) 
\,, \nonumber\\
\psi_{-q}^+(x,\bfk) &=& -\frac{k^1 + ik^2}{x} \, 
(1-x) \, \mu_q \, \varphi_q(x,\bfk) \,, \nonumber\\
&&\\
\psi_{+q}^-(x,\bfk) &=& \frac{k^1 - ik^2}{x} 
\, (1-x) \, \mu_q \, \varphi_q(x,\bfk) 
\,, \nonumber\\
\psi_{-q}^-(x,\bfk) &=& \frac{m_q + xM_N}{x} \, \varphi_q(x,\bfk) 
\,, \nonumber 
\en 
where $m_q = m_{1q}$ is the mass of struck quark; $\varphi_q(x,\bfk)$ 
is the wave function, which is related to 
the proposed LFWF~(\ref{Eff_WF}) as 
\eq\label{LFWF_fin} 
\varphi_q(x,\bfk) &=& N_q \, \sqrt{\log(1/x)} \, x^{\beta_{1q}} \, 
(1-x)^{\beta_{2q} + 1} 
\nonumber\\[1mm]
&\times& \exp\biggl[- \frac{{\cal M}^2}{2\kappa^2} \, x \, \log(1/x) \biggr] \, .
\en  
Here we include the flavor dependence effects in the normalization constant 
\eq 
N_q &=& \frac{4 \pi\sqrt{n_q}}{\kappa M_N}
\biggl[ \int\limits_0^1 dx \, x^{2\beta_{1q}}
(1-x)^{3+2\beta_{2q}} \nonumber\\
&\times&
R_q(x) \, e^{-\frac{{\cal M}_0^2}{\kappa^2} \, x \, \log(1/x)} 
\biggr]^{-1/2}   \,, \nonumber\\
\hspace*{-.6cm}
R_q(x) &=& \Big(1 + \frac{m_{1q}}{x M_N}\Big)^2
+  \frac{\kappa^2 \mu_q^2}{M_N^2} \frac{(1-x)^3}{\log(1/x) \, x^2}
\en
and in the parameters $\beta_{1q}$ and $\beta_{2q}$. 
The parameters $\mu_q$ $(q=u,d)$ are fixed  from the description 
of the nucleon magnetic moments. 
In the LFWF $\psi_{\mp q}^{\pm}(x,\bfk)$ 
we included an extra factor $(1-x)$ in order to generate 
an extra power $(1-x)^2$ in the helicity-flip parton 
density ${\cal E}^q(x) \sim (1-x)^5$ in comparison with 
the helicity non-flip density $q_v(x) \sim (1-x)^3$. Notice that
the extra power of $(1-x)$ in the functions $\psi_{\mp q}^{\pm}(x,\bfk)$ 
is consistent with results of the soft-wall AdS/QCD, 
where the functions  $\psi_{\mp q}^{\pm}(x,\bfk)$ and 
$\psi_{\pm q}^{\pm}(x,\bfk)$ have holographic analogues --- 
the bulk profiles of left- and right-handed AdS fermion fields 
with spin~$\frac{1}{2}$~\cite{Abidin:2009hr}-\cite{Gutsche:2011vb}.  

Substituting our ansatz for the LFWF~(\ref{LFWF_fin}) in Eq.~(\ref{FF_LFQCD}) 
and integrating over $\bfk$ we get the following expressions 
for the quark form factors: 
\eq
\hspace*{-.3cm}
F_1^q(Q^2) &=& C_q \, \int\limits_0^1 dx \,
x^{2\beta_{1q}} \, (1 - x)^{3+2\beta_{2q}} \, R_q(x,Q^2)
\nonumber\\
\hspace*{-.3cm}
&\times&
\exp\biggl[-\frac{Q^2}{4\kappa^2} \, \log(1/x) \, (1-x)\biggr]
\nonumber\\
\hspace*{-.3cm}
&\times&
\exp\biggl[ - \frac{{\cal M}_0^2}{\kappa^2} \, x \, \log(1/x) \biggr]  \,,
\nonumber\\
\hspace*{-.3cm}
& &\\
\hspace*{-.3cm}
F_2^q(Q^2) &=& C_q \int\limits_0^1 \frac{2dx}{x} \, \mu_q \, 
\biggl(1 + \frac{m_{1q}}{x M_N} \biggr) \, 
x^{2\beta_{1q}} \, (1 - x)^{5+2\beta_{2q}} \,  \nonumber\\
\hspace*{-.3cm}
&\times& 
\exp\biggl[-\frac{Q^2}{4\kappa^2} \, \log(1/x) \, (1-x)\biggr] 
\nonumber\\
\hspace*{-.3cm}
&\times& 
\exp\biggl[ - \frac{{\cal M}_0^2}{\kappa^2} \, x \, \log(1/x) \biggr]  \,, 
\nonumber
\en  
where 
\eq 
C_q &=& N_q^2 \ \Big(\frac{\kappa M_N}{4 \pi}\Big)^2  \,, \nonumber\\
R_q(x,Q^2) &=& R_q(x)
           - \frac{Q^2}{4M_N^2} \frac{(1-x)^4}{x^2} \, \mu_q^2\,.
\en 
In particular, one possible set of parameters, 
which gives a reasonable description 
of nucleon electromagnetic propertie is 
$m_1 = 7$ MeV (mass of active quark), $m_2 = 190$~MeV (mass of spectator 
diquark), $\mu_u = 0.3$, $\mu_d = -1.2$, $\beta_{1u} = 1.1$, 
$\beta_{2u} = 0.2$, 
$\beta_{1d} = 1.5$, 
$\beta_{2d} = 0.26$, 
$\kappa = 300$ MeV (see our results 
in comparison with data~\cite{Beringer:1900zz} in Table~I).   

\begin{table}[htb]
\begin{center}
\caption{Electromagnetic properties of nucleons} 

\vspace*{.2cm}

\def\arraystretch{1.5}
    \begin{tabular}{|c|c|c|}
      \hline
Quantity & Our results & Data~\cite{Beringer:1900zz}            \\
\hline
$\mu_p$ (in n.m.)          &  2.793       &  2.793              \\
\hline
$\mu_n$ (in n.m.)          & -1.826       & -1.913              \\
\hline
$r_E^p$ (fm)     &  0.741 &  0.8768 $\pm$ 0.0069 \\ 
\hline
$\la r^2_E \ra^n$ (fm$^2$) & -0.116 & -0.1161 $\pm$ 0.0022 \\ 
\hline
$r_M^p$ (fm)     &  0.817 &  0.777  $\pm$ 0.013 $\pm$ 0.010 \\ 
\hline
$r_M^n$ (fm)     &  0.851 &  0.862$^{+0.009}_{-0.008}$     \\
\hline
\end{tabular}
\end{center}
\end{table}

Now we demonstrate that the obtained results obey the 
model-independent scaling laws. Considering the limit 
$x \to 1$ where $\log(1/x) \sim 1-x$ and the limit 
$Q^2 \to \infty$ [changing $x \to 1-x$ and then rescaling  
$x \to (2 \kappa/\sqrt{Q^2}) \, x$], we prove the DYW duality 
and quark scaling rules  
\eq 
F_1^q(Q^2) &\sim& \frac{1}{(Q^2)^{2+\beta_{2q}}} \,, \nonumber\\
F_2^q(Q^2) &\sim& \frac{1}{(Q^2)^{3+\beta_{2q}}} \,, \\
\frac{F_2^q(Q^2)}{F_1^q(Q^2)} &\sim& \frac{1}{Q^2} 
\nonumber 
\en 
and 
\eq 
q_v(x) \sim (1-x)^{3+2\beta_{2q}} \,, \quad 
{\cal E}^q(x) \sim (1-x)^{5+2\beta_{2q}} \,, 
\en 
where $\beta_{2q} \to 0$ in the chiral limit. Therefore, 
the nucleon form factors and parton densities are consistent 
with the DYW duality and quark counting rules. We plan to present  
a detailed numerical analysis of GPDs and parton densities 
in a forthcoming paper~\cite{Paper_GPDs}. 

\section{Conclusion} 

In conclusion we want to stress again the main result of our paper: 
we demonstrated how to construct a light-front quark model 
consistent with the model-independent scaling laws --- 
the DYW duality~\cite{Drell:1969km} and 
quark counting rules~\cite{Brodsky:1973kr}.  
We plan to perform a detailed analysis of baryon properties 
using the light-front quark model proposed in the present letter. 

\begin{acknowledgments}

The authors thank Stan Brodsky and Guy de T\'e\-ra\-mond 
for useful discussions. 
This work was supported by the DFG under Contract No. LY 114/2-1, 
by FONDECYT (Chile) under Grant No. 1100287 and by CONICYT (Chile) 
under Grant No. 7912010025. The work is done partially under 
the project 2.3684.2011 of Tomsk State University. 
V. E. L. would like to thank Departamento de F\'\i sica y Centro
Cient\'\i fico Tecnol\'ogico de Valpara\'\i so (CCTVal), Universidad
T\'ecnica Federico Santa Mar\'\i a, Valpara\'\i so, Chile for warm
hospitality. 

\end{acknowledgments}

\end{document}